\documentstyle[11pt,newpasp,twoside,epsf]{article}
\def\edcomment#1{\iffalse\marginpar{\raggedright\sl#1\/}\else\relax\fi}
\reversemarginpar

\begin{document}
\title{The DIRECT Project}
\author{Lucas M. Macri}
\affil{Hubble Fellow, National Optical Astronomy Observatory, 950 N Cherry Avenue, Tucson,
 AZ 85719, United States of America}

\begin{abstract}
The DIRECT project aims to determine direct distances to two important galaxies
in the cosmological distance ladder -- M31 and M33 -- using detached eclipsing
binaries (DEBs) and Cepheids. I present an overview of the project and some results from follow-up observations.
\end{abstract}

\section{Introduction}

The DIRECT project (as in ``direct distances'') started in 1996 with the
long-term goal of obtaining distances to two important galaxies in the
cosmological distance ladder -- M31 and M33 -- using detached eclipsing
binaries (DEBs) and Cepheids.  These two nearby galaxies are the stepping
stones in most of the current effort to understand the evolving universe at
large scales. Not only are they essential to the calibration of the
extragalactic distance scale, but they also constrain population synthesis
models for early galaxy formation and evolution. However, accurate distances
are essential to make these calibrations free from large systematic
uncertainties.

Detached eclipsing binaries have the potential to establish distances to M31
and M33 with an unprecedented accuracy of better than 5\% and possibly to
better than 1\%. Current uncertainties in the distances to these galaxies are
in the order of 10 to 15\%, as there are discrepancies of 0.2-0.3 mag between
various distance indicators. Detached eclipsing binaries (Paczy\'nski 1997)
offer a single-step distance determination to nearby galaxies (Fitzpatrick et
al. 2003) and may therefore provide an accurate zeropoint calibration for other
distance indicators, including Cepheids.

\section{DIRECT Observations and Results}

The DIRECT project obtained time-series observations of M31 and M33 during 170
nights between 1996 September 6 and 2000 January 2 using the Fred L. Whipple
Observatory 1.2-m telescope. Additional observations were carried out during 36
nights in 1996 and 1997 at the Michigan-Dartmouth-MIT Observatory 1.3-m
telescope. In 1996 and 1997, images were obtained using a camera with $\sim
11\arcmin \times 11\arcmin$ field of view. In 1998 and 1999, data were
collected using a camera with a $\sim 22\arcmin \times 22\arcmin$ field of
view.Observations were obtained mainly in the $V$ and $I$ bands, with some
additional data in the $B$ band. The total area covered by the observations was
$\sim 0.5\sq \deg$ in M31 and $\sim 0.3\sq \deg$ in M33 (see Figure 1).

\begin{figure}
\plottwo{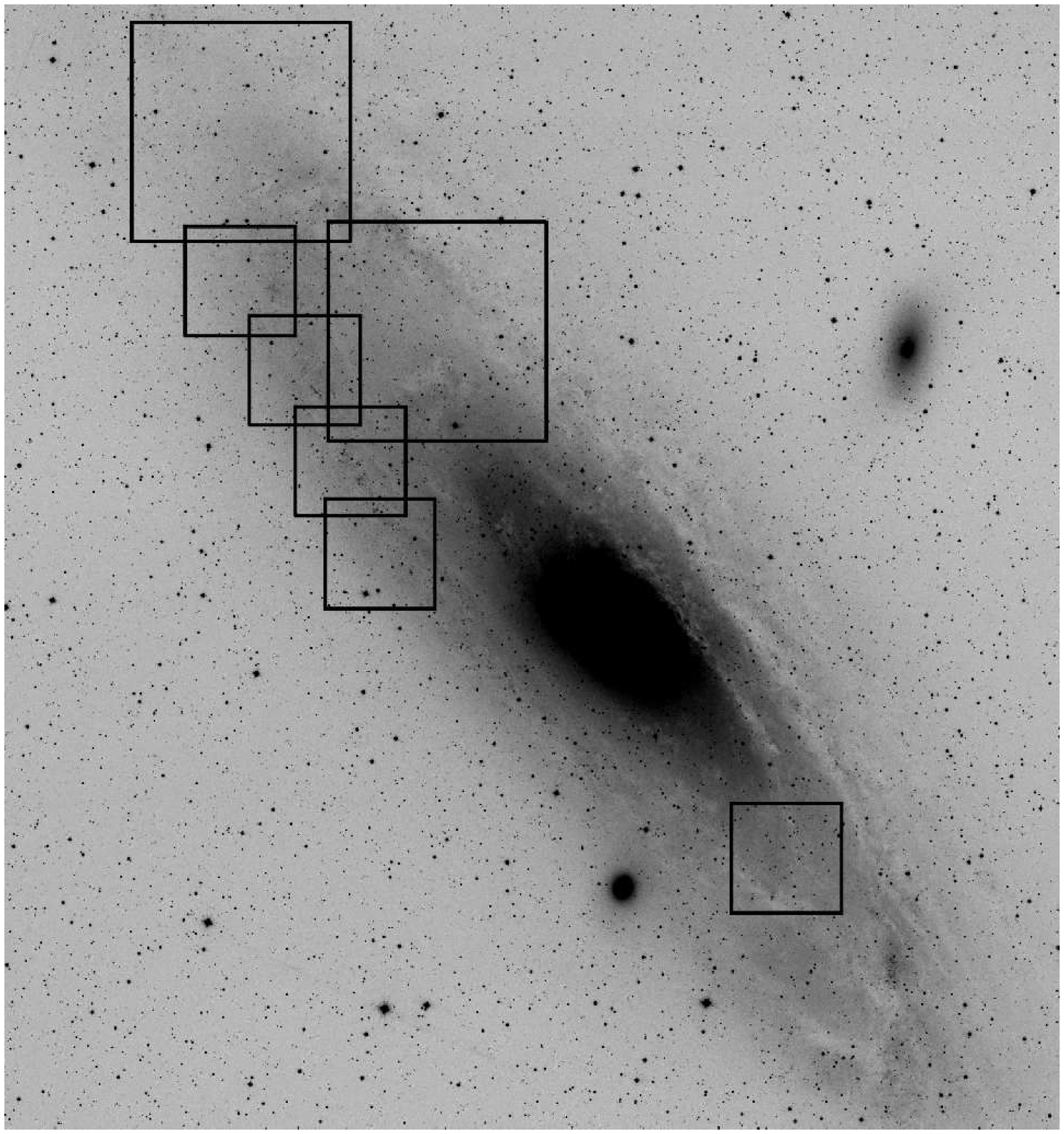}{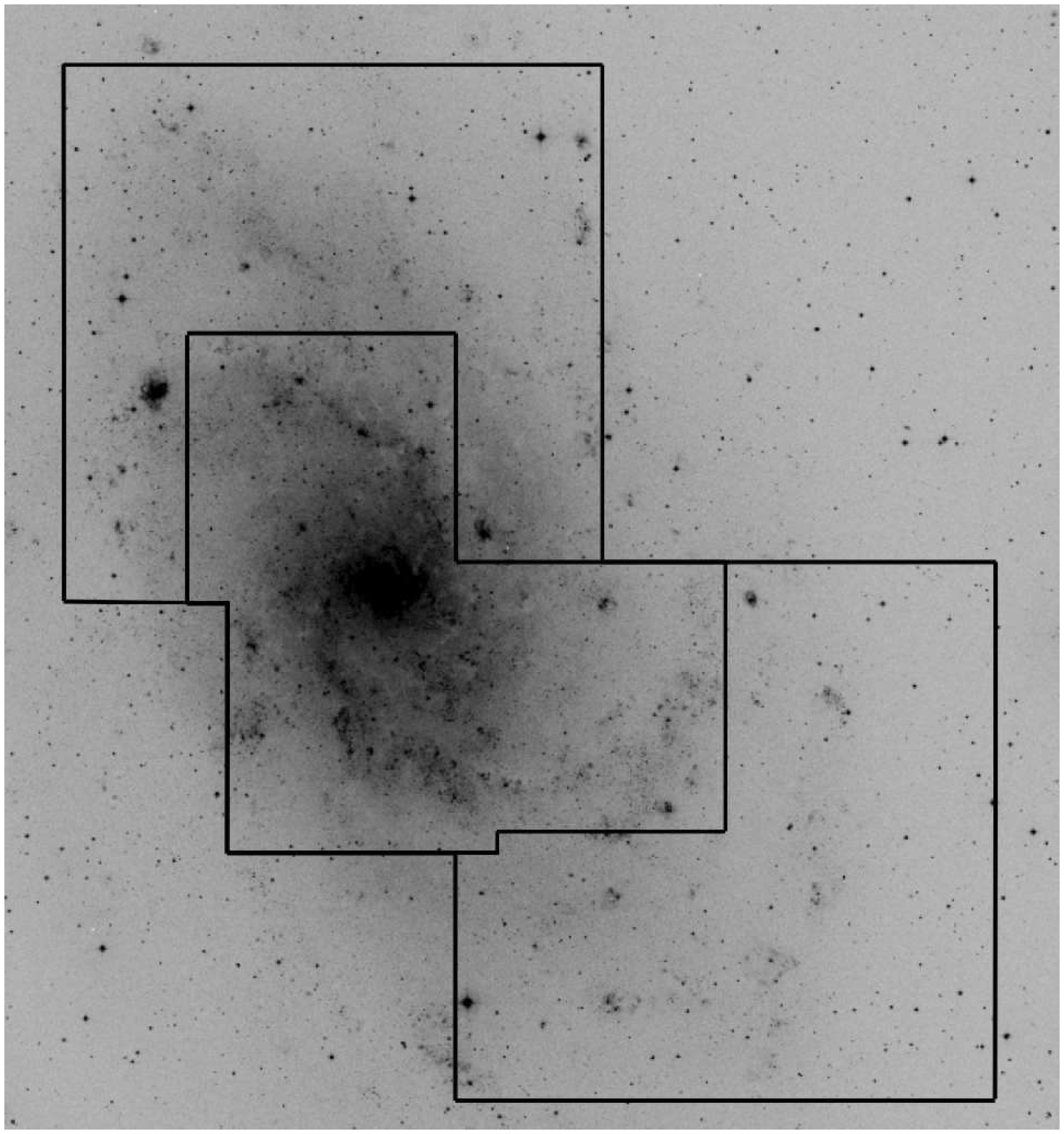}
\caption{DIRECT fields in M31 and M33. The small and large fields are
$11\arcmin$ and $22\arcmin$ on a side, respectively. M31 image: Schoening \&
Harvey/REU/NOAO/AURA/NSF. M33 image: DPOSS II \copyright\ CalTech.}
\end{figure}

Most of the fields have been analyzed and published (Bonanos et al. 2003;
Kaluzny et al. 1998; Kaluzny et al. 1999; Macri et al. 2001a; Mochejska et
al. 1999; Stanek et al. 1998; Stanek et al. 1999). These publications contain a
total of $\sim 130$ eclipsing binaries, $\sim 600$ Cepheid variables, and $\sim
500$ miscellaneous variables. Representative light curves of Cepheid variables
and eclipsing binaries discovered in M33 can be seen in Figures 2 and 3. Two
additional publications will appear in the near future with the analysis of the
remaining fields (Bonanos et al. 2004; Macri et al. 2004).

Additional publications related to the DIRECT project include: the discovery of
cessation of pulsations in a long-period Cepheid in M33 (Macri, Sasselov, \&
Stanek 2001); the study of the influence of unresolved blends on the Cepheid
Distance Scale (Mochejska et al. 2000; Mochejska et al. 2001a); a catalog of 
globular clusters in M31 and M33 (Mochejska et al. 1998); ancillary
stellar catalogs and additional short-period variables (Macri et al. 2001b;
Mochejska et al. 2001b; Mochejska et al. 2001c; Mochejska et al. 2001d).

\begin{figure}
\plotfiddle{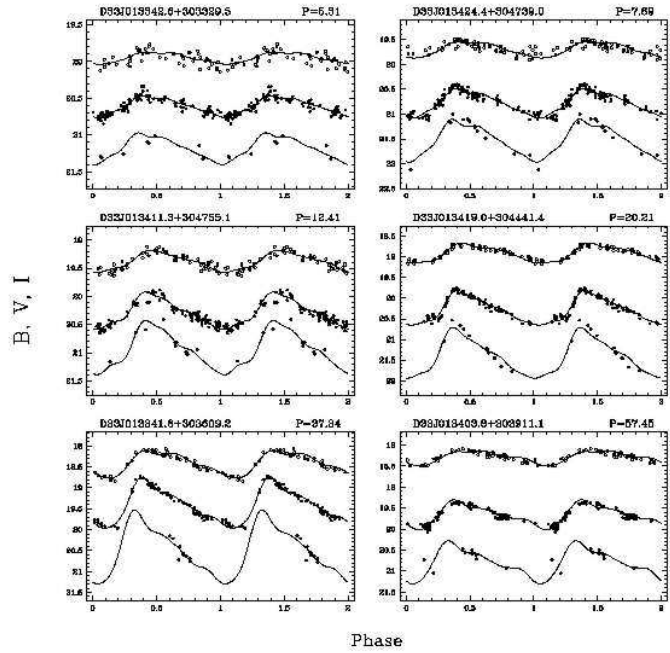}{3in}{0}{47}{47}{-144}{-15}
\caption{Representative light curves of Cepheid variables discovered by DIRECT in M33 (Macri et al. 2001a).}
\end{figure}

\begin{figure}
\plotfiddle{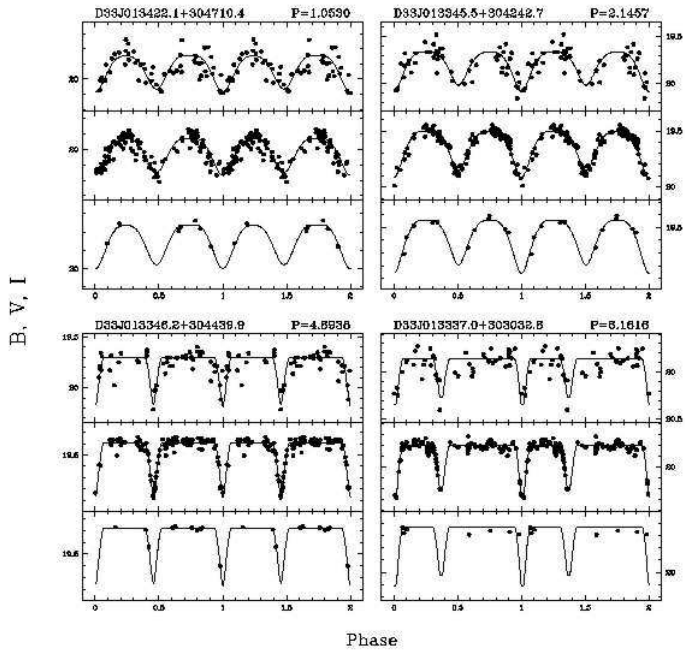}{3in}{0}{47}{47}{-144}{-15}
\caption{Representative light curves of eclipsing binaries discovered by DIRECT in M33 (Macri et al. 2001a).}
\end{figure}

\section{Follow-up observations}
\subsection{Detached eclipsing binaries}

Four promising detached eclipsing binary systems have been discovered in the
DIRECT fields; two in M31 and two in M33. As a first step in their follow-up,
higher quality light curves were obtained through observations
conducted at the Kitt Peak National Observatory 2.1-m telescope in 1999 and
2001. ecently, a program was started to measure radial velocities of
the DEBs using the Echelle Spectrograph and Imager (ESI) at Keck Observatory.
Figure 4 shows the radial velocities measured for a system in M33 during the
first two nights of that program, plotted against the radial velocity predicted
by the photometric data. 

\begin{figure}[hbt]
\plotfiddle{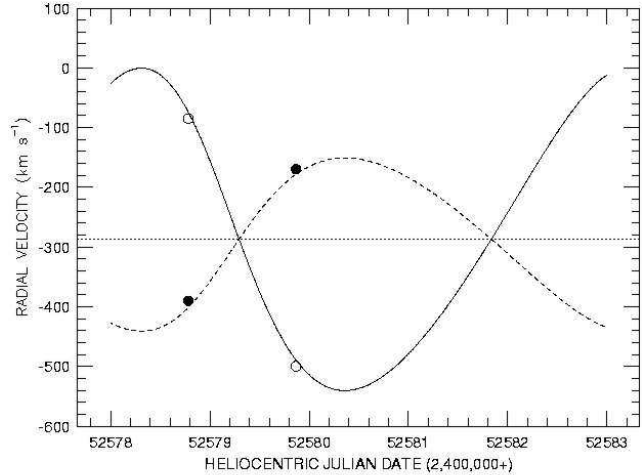}{2.5in}{0}{55}{55}{-170}{-40}
\caption{Radial velocities for one detached eclipsing binary system discovered
by DIRECT, obtained during two nights of observation with ESI on Keck. The
solid and dashed lines show the variation with phase predicted by the
photometric data.}
\end{figure}

\subsection{Cepheid variables}

M33 has one of the largest abundance gradients among nearby spiral galaxies,
$\sim 0.2$~dex/kpc (Henry \& Howard 1995; Monteverde, Herrero, \& Lennon
2000). The DIRECT Cepheid sample in this galaxy exceeds 700 variables and
covers the majority of the disk, making it ideal for a study of the
metallicity dependence of the Cepheid Period-Luminosity relation. A previous 
observational determination (Kennicutt et al. 1998) based on a similar differential test, but using Cepheids in M101, obtained a value for the dependence in the $VI$ bands that is at odds with theoretical predictions (Bono et al. 1999). However, a recent analysis by Fiorentino et al. (2002) has postulated a possible solution for the disagreement.

Follow-up observations of M33 Cepheids were started in 2002 August using the
MiniMosaic CCD imager at the WIYN 3.5-m telescope. Once these observations
conclude in 2004 January, we expect to have obtained $\sim 15$~epochs in $BVI$
with a greatly improved image quality relative to the original DIRECT survey
($0.7\arcsec$ vs. $1.5\arcsec$), which in turn will result in greatly improved
photometric accuracy. We should be able to detect the metallicity effect
predicted by Fiorentino et al. (2002) in the $BVI$ bands at the $6\sigma$
level.

Additionally, we are carrying out near-infrared ($JHK_s$) observations, using
the Near InfraRed Imager (NIRI) on the 8.1-m Fred Gillett Gemini North
telescope, of a subsample of $\sim 200$ Cepheids well distributed across the
disk of M33. Figure 5 shows preliminary $H$ and $K_s$ P-L relations for a
subset of 95 variables located in the inner part of the galaxy. Once all
data are collected, we should be able to detect the metallicity effect
predicted by Fiorentino et al. (2002) in the $K_s$-band at the $4\sigma$ level.

\begin{figure}[hbt]
\plotfiddle{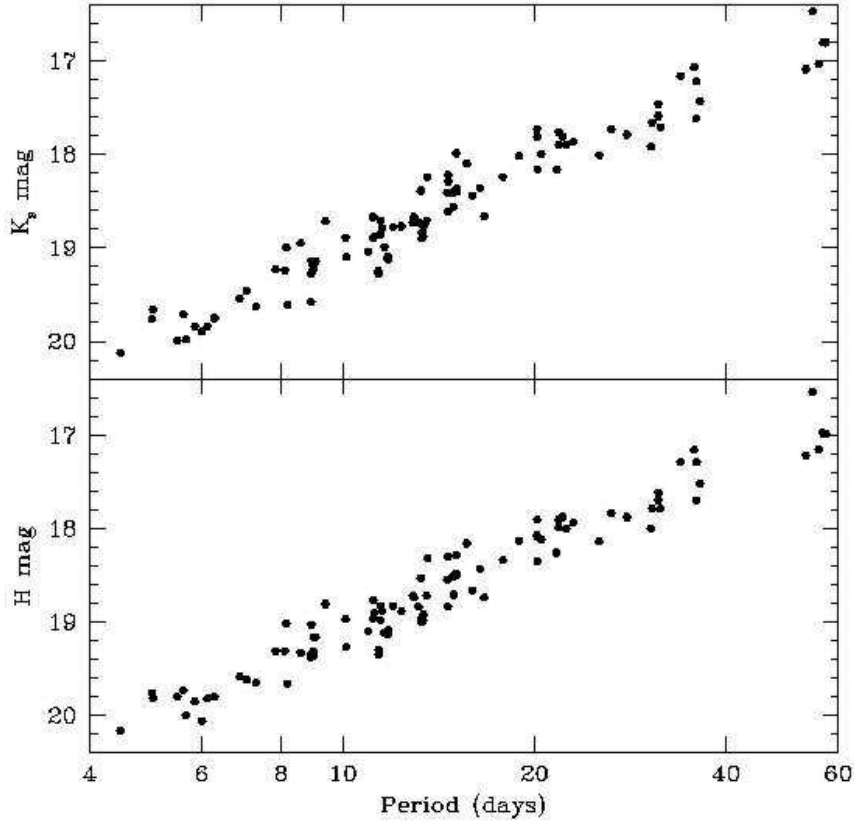}{4.25in}{0}{60}{60}{-180}{-10}
\caption{$K_s$ and $H$-band Period-Luminosity relations for a subsample of 95
Cepheids located in the central part of M33. Based on observations obtained
with NIRI on Gemini North.}
\end{figure}

\acknowledgments

Support for this work was provided by NASA through Hubble Fellowship
grant \#HST-HF-01153.01 awarded by the Space Telescope Science
Institute, which is operated by the Association of Universities for
Research in Astronomy, Inc., for NASA, under contract NAS 5-26555.


\begin{references}
Bonanos, A.Z., Stanek, K.Z., Sasselov, D.D., Mochejska, B.J., Macri, L.M., \& Kaluzny, J. 2003, \aj, 126, 175

Bonanos, A.Z., Stanek, K.Z., Sasselov, D.D., Mochejska, B.J., Macri, L.M., \& Kaluzny, J. 2004, in preparation

Bono, G., Caputo, F., Castellani, V., \& Marconi, M. 1999, \apj, 512, 711

Fiorentino, G., Caputo, F., Marconi, M., Musella, I. 2002, \apj, 576, 402

Fitzpatrick, E.L., Ribas, I., Guinan, E.F., Maloney, F.P., \& Claret, A. 2003, ApJ, 587, 685

Henry, R.B.C. \& Howard, J.W. 1995, \apj, 438, 170

Kaluzny, J., Stanek, K.Z., Krockenberger, M., Sasselov, D.D., Tonry, J.L., \& Mateo, M. 1998, \aj, 115, 1016

Kaluzny, J., Mochejska, B.J., Stanek, K.Z., Krockenberger, M., Sasselov, D.D., Tonry, J.L., \& Mateo, M.~1999, \aj, 188, 346

Kennicutt, R.C., Stetson, P.B., Saha, A., Kelson, D., Rawson, D.M., Sakai, S., Madore, B.M., Mould J.R., Freedman, W.L., Bresolin, F., Ferrarese, L., Ford, H., Gibson, B.K., Graham, J.A., Han, M., Harding, P., Hoessel, J.G., Huchra, J.P., Hughes, S.M.G, Illingworth, G.D., Macri, L.M., Phelps, R.L., Silbermann, N.A., Turner, A.M., \& Wood, P.R. 1998, \apj, 498, 181

Macri, L.M., Stanek, K.Z., Sasselov, D.D., Krockenberger, M., \& Kaluzny, J. 2001a, \aj, 120, 870

Macri, L.M., Stanek, K.Z., Sasselov, D.D., Krockenberger, M., \& Kaluzny, J. 2001b, \aj, 120, 861

Macri, L.M, Sasselov, D.D. \& Stanek, K.Z. 2001, \apj, 550, L159

Macri, L.M., Stanek, K.Z., Sasselov, D.D., Mochejska, B.J., Bonanos, A.Z., \& Kaluzny, J. 2004, in preparation

Mochejska, B.J., Kaluzny, J., Krockenberger, M., Sasselov, D.D., \& Stanek, K.Z. 1998, Ac.A., 48, 445

Mochejska, B.J., Kaluzny, J., Stanek, K.Z., Krockenberger, M., \& Sasselov, D.D.~1999, \aj, 118, 2211

Mochejska, B.J., Macri, L.M., Sasselov, D.D., \& Stanek, K.Z. 2000, \aj, 120, 810

Mochejska, B.J., Macri, L.M., Sasselov, D.D., \& Stanek, K.Z. 2001a, astro-ph/0103440

Mochejska, B.J., Kaluzny, J., Stanek, K.Z., \& Sasselov, D.D. 2001b, \aj, 122, 1383

Mochejska, B.J., Kaluzny, J., Stanek, K.Z., Sasselov, D.D., \& Szentgyorgyi, A.H. 2001c, \aj, 121, 2032

Mochejska, B.J., Kaluzny, J., Stanek, K.Z., Sasselov, D.D., \& Szentgyorgyi, A.H. 2001d, \aj, 122, 2477

Monteverde, M.I., Herrero, A., \& Lennon, D.J. 2000, \apj, 545, 813

Paczy\'nski, B. 1997, in The Extragalactic Distance Scale, ed. M. Livio, M. Donahue \& N. Panagia (Cambridge: Cambridge Univ. Press), 273

Stanek, K.Z., Kaluzny, J., Krockenberger, M., Sasselov, D.D., Tonry, J.L., \& Mateo, M. 1998, \aj, 115, 1894

Stanek, K.Z., Kaluzny, J., Krockenberger, M., Sasselov, D.D., Tonry, J.L., \& Mateo, M.~1999, \aj, 117, 2810


\end{references}
\end{document}